\documentclass{ws-ijmpc_rev}
\usepackage{graphicx}

\begin{document}

\markboth{Hiroki Mizuno and Hiroshi Koibuchi}
{In-plane deformation of a triangulated surface model}

\catchline{}{}{}{}{}

\title{In-plane deformation of a triangulated surface model with metric degrees of freedom
}

\author{Hiroki Mizuno and Hiroshi Koibuchi
}

\address{Department of Mechanical and Systems Engineering, Ibaraki National College of Technology, Nakane 866 Hitachinaka, Ibaraki 312-8508, Japan
\\ koibuchi@mech.ibaraki-ct.ac.jp}



\maketitle

\begin{history}
\end{history}

\begin{abstract}
Using the canonical Monte Carlo simulation technique, we study a Regge calculus model on triangulated spherical surfaces. The discrete model is statistical mechanically defined with the variables $X$, $g$ and $\rho$, which denote the surface position in ${\bf R}^3$, the metric on a two-dimensional surface $M$ and the surface density of $M$, respectively.  The metric $g$ is defined only by using the deficit angle of the triangles in {$M$}. This is in sharp contrast to the conventional Regge calculus model, where {$g$} depends only on the edge length of the triangles. We find that the discrete model in this paper undergoes a phase transition between the smooth spherical phase at $b\!\to\!\infty$ and the crumpled phase at $b\!\to\!0$, where $b$ is the bending rigidity. The transition is of first-order and identified with the one observed in the conventional model without the variables $g$ and $\rho$. This implies that the shape transformation transition is not influenced by the metric degrees of freedom. It is also found that the model undergoes a continuous transition of in-plane deformation. This continuous transition is reflected in almost discontinuous changes of the surface area of $M$ and that of $X(M)$, where the surface area of $M$ is conjugate to the density variable $\rho$. 

\keywords{Triangulated surface model; Regge calculus model; Monte Carlo; First-order transition; In-plane transition}
\end{abstract}

\ccode{PACS Nos.: 11.25.-w,  64.60.-i, 68.60.-p, 87.10.-e, 87.15.ak}

\section{Introduction}
A surface model for membranes is defined by a mapping $X$ from a two-dimensional surface $M$ to ${\bf R}^3$ \cite{NELSON-SMMS2004,Bowick-PREP2001,WIESE-PTCP19-2000,GOMPPER-KROLL-SMMS2004}. Not only the mapping $X$ but also the metric $g$ of $M$ is assumed as dynamical variables of the model in its statistical mechanical study \cite{WHEATER-JP1994}. However, the variable $g$ is always assumed to be the Euclidean metric $g_{ab}\!=\!\delta_{ab}$ or the induced metric $g_{ab}\!=\!\partial_a X^\mu \partial_b X^\mu$ of the mapping $X$ in the numerical studies that have been performed so far \cite{KANTOR-NELSON-PRA1987}. In those studies, the variable $g$ is fixed in the partition function, and as a consequence a role of $g$ in the phase structure of the model remains unclear. 

The metric $g$ of $M$ can be discretized as a variable by using the edge length of triangles in the Regge calculus approach \cite{REGGE-NC1961,HAMBER-LH1986,FDAVID-LH1992} to the triangulated surface model. The model is defined by the Hamiltonian $S\!=\!S_1+bS_2$, where $S_1$ and $S_2$ are the Gaussian bond potential and the bending energy, and $b[kT]$ is the bending rigidity \cite{HELFRICH-1973,POLYAKOV-NPB1986,KLEINERT-PLB1986}. In this approach, one can see whether or not $g$ influences the discontinuous transition between the smooth phase at $b\!\to\!\infty$ and the collapsed phase at $b\!\to\! 0$. This transition and related phenomena have long been studied numerically as well as analytically by many groups \cite{CATTERALL-NPBSUP1991,AMBJORN-NPB1993,Munkel-Heermann-JPF1994,Munkel-Heermann-PRL1995,P-L-1985PRL,David-1986EPL,DavidGuitter-1988EPL,BKS-2000PLA,BK-2001EPL,KD-PRE2002,Kownacki-Mouhanna-2009PRE,Essafi-Kownacki-Mouhanna-2011PRL,NISHIYAMA-PRE-2004,SWAMM-PRL-2010,Hasselmann-Braghin-PRE-2011}.

Such a Regge calculus model was recently studied in Ref. \refcite{KOIB-NPB2010}, where $M$ is embedded in ${\bf R}^2$ and the metric $g_{ab}$ is slightly extended from the conventional Regge metric by incorporating a degree of freedom for a deficit angle of the triangles in $M$. One of the non-trivial results reported in Ref. \refcite{KOIB-NPB2010} is that the deficit angle  varies almost discontinuously at the transition point, although the discontinuity is very small compared to the deficit angle itself. 

However, it is still unclear whether or not such a discontinuity in the internal geometric quantities can also be seen in the case where the edge length of triangles is fixed while the deficit angle is varied. Moreover, it is also  nontrivial  whether  the fluctuating metric influences the phase structure of the model with a constant metric.

In this paper, we study a surface model with a metric variable, which depends only on the deficit angle of triangles. Since the surface is embedded in ${\bf R}^3$, the numerical simulation is more time consuming than that of the model in Ref. \refcite{KOIB-NPB2010}. However, the metric is simplified by fixing the edge length $L$ of the triangle in $M$ such that $L\!=\!1$, and hence the numerical simulations are greatly simplified compared to those for the model in Ref. \refcite{KOIB-NPB2010}. 

We should note that the surface area of $X(M)$ varies if the parameter $a$ in $S(X)\!=\!aS_1\!+\!bS_2$ is varied while $b$ is fixed. However, the parameter $a$ can always be fixed to $a\!=\!1$  because of the scale transformation  $X\!\to\! X^\prime\!=\!(1/\sqrt{a})X$. Indeed, $S(X)$ changes from $S(X)\!=\!aS_1\!+\!bS_2$ to $S(X)\!=\!S_1\!+\!bS_2$ while the partition function $Z$ remains unchanged; it changes up to a multiplicative constant, under this scale transformation. This implies that $a$ can always be fixed to $a\!=\!1$. 

\section{Discrete surface model}
A discrete Hamiltonian and the partition function are introduced in this section. These discrete quantities are closely related to those in Ref. \refcite{KOIB-NPB2010}, where the two-dimensional surface $M$ is embedded in ${\bf R}^2$ by a mapping $X$, while in this paper $M$ is embedded in ${\bf R}^3$. 

\begin{figure}[hbt]
\centering
\includegraphics[width=8.5cm]{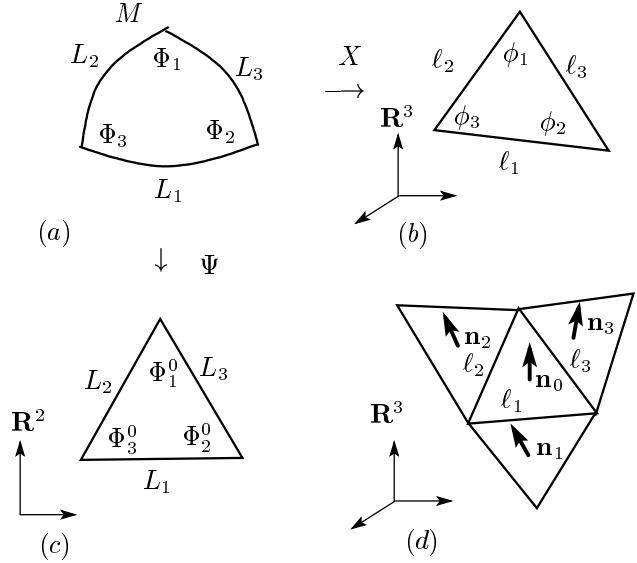}
\caption{(a) The smooth triangle ${\it \Delta}$ in $M$,  (b) the image $X({\it \Delta})$ of  ${\it \Delta}$ by a mapping $X$ from $M$ to the external space ${\bf R}^3$, (c) a local coordinate domain $D$ in  ${\bf R}^2$ and the regular triangle $\Psi({\it \Delta})$, and (d) a unit normal vector ${\bf n}_0$ of $X({\it \Delta})$ and those ${\bf n}_i(i\!=\!1,2,3)$ of the nearest neighbor triangles.}
\label{fig-1}
\end{figure}
A two-dimensional surface $M$ is smoothly triangulated, and a triangle ${\it \Delta}$ is shown in Fig. \ref{fig-1}(a), where $L_i(i\!=\!1,2,3)$ denotes the edge length, and $\Phi_i(i\!=\!1,2,3)$ denotes the internal angle of  ${\it \Delta}$. The triangle ${\it \Delta}$ in $M$ is mapped into ${\bf R}^3$ by $X$ as shown in Fig. \ref{fig-1}(b), where the triangle $X({\it \Delta})$ is a linear one. The edge lengths and the internal angles of $X({\it \Delta})$ are denoted by $\ell_i$ and $\phi_i$, respectively.  

Let $\Psi$ denote a coordinate mapping from ${\it \Delta}$ to a domain $D$ in ${\bf R}^2$, then we have a linear triangle  $\Psi({\it \Delta})$ in $D$, which is shown in Fig. \ref{fig-1}(c). The edge length $L_i(i\!=\!1,2,3)$ of ${\it \Delta}$ is assumed to be identical to the one in $\Psi({\it \Delta})$, while the internal angle $\Phi_i(i\!=\!1,2,3)$ of ${\it \Delta}$ differs from $\Phi_i^0(i\!=\!1,2,3)$ of $\Psi({\it \Delta})$, where $\sum_{i=1}^3 \Phi^0_i\!=\!2\pi$. In the conventional Regge calculus model, there is no difference between ${\it \Delta}$ in Fig. \ref{fig-1}(a) and  $\Psi({\it \Delta})$ in Fig. \ref{fig-1}(c) because $\Phi_i=\Phi^0_i$. 

As mentioned in the Introduction, one basic assumption is that  
\begin{equation} 
\label{triangle-rel-1}
 L_1=L_2=L_3=1 
\end{equation} 
for all ${\it \Delta}$. This implies that $\Phi_i^0\!=\!\pi/3(i\!=\!1,2,3)$. Another assumption is expressed as
\begin{equation} 
\label{triangle-rel-2}
 \Phi_1=\Phi_2=\Phi_3=\Phi, 
\end{equation} 
where $\Phi$ depends on  ${\it \Delta}$, and $\Phi$ is not always identical to $\pi/3$ as in the model of Ref. \refcite{KOIB-NPB2010}. We should note that "smoothly triangulated" does not always mean $\sum_{j(i)} \Phi_{j(i)}\!=\!2\pi$, where $\Phi_{j(i)}$ is an internal angle meeting at the vertex $i$ of ${\it \Delta}$ in $M$ \cite{KOIB-NPB2010}. 

Since $\Phi$ is not always given by $\Phi\!=\!\pi/3$, the deficit angle $\varphi$ can be defined such that
\begin{equation} 
\label{deficita}
\varphi=3\Phi-\pi,
\end{equation}
or equivallently
\begin{equation} 
\label{internal_angle}
 \Phi=\Phi^0\left(1+\frac {\varphi}{\pi}\right)={\frac 1 3}\left(\pi+\varphi\right). 
\end{equation} 
We should note that ${\it \Delta}$ is not always a linear triangle, because the edges of ${\it \Delta}$ are curved and the deficit angle $\varphi$ is not always zero. On the contrary, the triangle $X({\it \Delta})$ is linear, because  the deficit angle of $X({\it \Delta})$ is fixed to be zero.

The surface model is defined by the discrete metric 
\begin{equation} 
\label{induced_metric}
g_{ab}=\left(  
       \begin{array}{@{\,}ll}
        1& \; F \\
       F & \; 1 
       \end{array} 
       \\ 
 \right), \quad  F=\cos \Phi,\quad |F|<1,
\end{equation} 
where $F$ represents the deficit angle $\varphi$. This $g_{ab}$ is the induced metric of the coordinate mapping $\Psi$.  The inequality in Eq. (\ref{induced_metric}) implies that $ds^2\!=\!\sum_{ab}g_{ab}dx_adx_b$ is positive definite. The area of ${\it \Delta}$ in $M$ is given by 
\begin{equation}
\label{area}
A_{\it \Delta}={\frac 1 2}\sqrt{1-F^2}.
\end{equation} 
 The area $A_{\it \Delta}$ varies according to the variation of $F$ although the edge length $L$ is fixed.  Since $g$ varies as a function of $F$, we call the model as a Regge calculus model even though $L$ is fixed.

The Hamiltonian is defined by 
\begin{eqnarray}
\label{Disc-Eneg}
&&S\left(X,g\right)=S_1+bS_2,  \\
&&S_1=\frac {1}{ 6}\sum_{\it \Delta} S_1\left({\it \Delta}\right)/A_{\it \Delta},\quad
S_2=\frac {1}{ 6}\sum_{\it \Delta} S_2\left({\it \Delta}\right)/A_{\it \Delta}, \nonumber
\end{eqnarray}
with
\begin{equation}
\label{Disc-Eneg-S1} 
S_1\left({\it \Delta}\right)=\ell_1^2+\ell_2^2+\ell_3^2  
-F_{\it \Delta}\left(\ell_1\ell_2\cos\phi_3 +\ell_2\ell_3\cos\phi_1+ \ell_3\ell_1\cos\phi_2\right)
\end{equation}
and
\begin{eqnarray}
\label{Disc-Eneg-S2} 
&&S_2\left({\it \Delta}\right)
= 3\!-\!{\bf n}_0\!\cdot\!{\bf n}_1\!-\!{\bf n}_0\!\cdot\!{\bf n}_2\!-\!{\bf n}_0\!\cdot\!{\bf n}_3  \nonumber \\
&&-F_{\it \Delta}[\left({\bf n}_1\!-\!{\bf n}_0\right)\!\cdot\!\left({\bf n}_2\!-\!{\bf n}_0\right) 
+\left({\bf n}_2\!-\!{\bf n}_0\right)\!\cdot\!\left({\bf n}_3\!-\!{\bf n}_0 \right)  
+\left({\bf n}_3\!-\!{\bf n}_0\right)\!\cdot\!\left({\bf n}_1\!-\!{\bf n}_0\right)],  
\end{eqnarray} 
where the symbol $\phi_i$ is an internal angle of $X({\it \Delta})$ as mentioned above. $F_{\it \Delta}$ in Eqs. (\ref{Disc-Eneg-S1}) and (\ref{Disc-Eneg-S2}) denotes that the value of $F\!=\!\cos \Phi$ depends on the triangle ${\it \Delta}$. The unit normal vectors ${\bf n}_i(i\!=\!0,1,2,3)$ are shown in Fig. \ref{fig-1}(d).

The partition function $Z(b)$ is defined by 
\begin{equation} 
\label{part_funct}
Z(b)=\int{\cal D}g \int {\cal D}X \exp\left[-S(X,g)\right].
\end{equation} 
 The integration $\int {\cal D}X$ in $Z(b)$ is given by the multiple $3$-dimensional ones such that
\begin{equation} 
\label{measure_X}
\int {\cal D}X = \int^\prime \prod _{i=1}^{N} d X_i, 
\end{equation} 
where $\int^\prime $ denotes that the center of mass of the surface is fixed to the origin of ${\bf R}^3$. The symbol $\int {\cal D}g$ in $Z(b)$ is defined by
\begin{eqnarray} 
\label{measure}
&&\int {\cal D}g=
\int\prod _{i=1}^{N_T} d F_i\int\prod _{i=1}^{N_T} d \rho_i\exp\left(-\lambda_F S_F-\lambda_{\rho A} S_{\rho A} -\lambda_\rho S_\rho\right), \nonumber \\
&&|F_i|<1,\quad \rho_i>0.
\end{eqnarray} 
 The integration $\int\prod _{i=1}^{N_T} d \rho_i$ is included in  $\int {\cal D}g$ although $\rho$ is a variable independent of the surface geometry. The symbol $N_T\!=\!2N\!-\!4$ in Eq. (\ref{measure}) is the total number of triangles. In the exponential factor in Eq. (\ref{measure}),  $S_F$ is the interaction term for the variable $F$ given by
\begin{equation}
\label{Disc-Eneg-Sg}
S_F=\sum_{ij} |F_i-F_j|,
\end{equation}
where $\sum_{ij}$ represents all nearest neighbor triangles $i$ and $j$. In the presence of the interaction $S_F$, the metric  $g_{ab}$ can be smoothened as a function on the surface $M$. Thus, it is physically natural to include $S_F$ as a measure term. We should note that the interaction $|F_i-F_j|$ in Eq. (\ref{Disc-Eneg-Sg}) can also be expressed by $(\partial F)^2$, however we use the expression $|F_i-F_j|$ for numerical simplicity.

 $S_{\rho A}$ and $S_{\rho}$ in Eq. (\ref{measure}) are given by  
\begin{equation}
\label{Disc-Eneg-rho} 
S_{\rho A}=\sum_{\it \Delta} \rho_{\it \Delta} A_{\it \Delta},\quad S_{\rho}=\sum_{ij} |\rho_i-\rho_j |,
\end{equation}
where $\rho_{\it \Delta}$ (or  $\rho_i$) is the conjugate variable to the area $A_{\it \Delta}$ (or  $A_i$) and can be called the "surface density". We should note that the surface density $\rho_{\it \Delta}$ is not an external field but introduced as a variable field on the surface to see the in-plane surface deformation. 

 The density $\rho_{\it \Delta}$ is interconnected to the geometric variables $X$ and $g$ only through the interaction $\rho_{\it \Delta} A_{\it \Delta}$. The term $S_\rho$ is introduced to define an interaction between the fileds $\rho_{\it \Delta}$. Since the variable $\rho$ couples to $A$ in $S_{\rho A}$, then $\rho$ becomes nonzero finite (or well-defined). We should note also that $A_{\it \Delta}$ can vary without the density field $\rho$ because $A_{\it \Delta}$ depends on $F$ which varies on the surface $M$. However, the interaction $S_{\rho A}$ is expected to influence the in-plane deformation of $A_{\it \Delta}$. 

By including the measure terms of the exponential factor of Eq. (\ref{measure}) in the Hamiltonian, we have the effective Hamiltonian such that 
\begin{equation} 
S\left(X,F,\rho\right)=S_1+bS_2+\lambda_F S_F+\lambda_{\rho A} S_{\rho A}+\lambda_\rho S_\rho.
\end{equation}
Note also that the conventional Hamiltonians  $S_1\!=\!\sum_{ij}(X_i\!-\!X_j)^2$ and $S_2\!=\!\sum_{ij}(1\!-\!{\bf n}_i\cdot{\bf n}_j)$  are restored up to irrelevant multiplicative factors if $F\!=\!0$. In this case, we have $S_F\!=\!0$ and $A_{\it \Delta}\!=\!1/2$, and therefore we find that both $S_{\rho A}$ and $S_\rho$ are independent of the surface and can be neglected.

\section{Continuous surface model}
\label{cont_model}
The surface model of Helfrich and Polyakov is defined by a mapping $X$ from a two-dimensional surface $M$ to ${\bf R}^3$, described by $X:M \ni  (x_1,x_2)\mapsto X(x_1,x_2)\in {\bf R}^3$. The surface $M$ is assumed to be of sphere topology in this paper just like in Ref. \refcite{KOIB-NPB2010}. The symbol $(x_1,x_2)$ denotes a local coordinate of $M$. The image $X(M)$ corresponds to a real physical membrane, however, the self-avoiding property is not assumed in $X(M)$, and hence the mapping $X$ is not always injective. 

The Hamiltonian of the model is given by a linear combination of the Gaussian bond potential $S_1$ and the extrinsic curvature energy $S_2$ such that
\begin{eqnarray} 
\label{cont_S}
&&S=S_1+bS_2,  \nonumber \\
&&S_1=\int \sqrt{g}d^2x g^{ab} \partial_a X^\mu \partial_b X^\mu, \\
&&S_2=\frac{1}{2}\int \sqrt{g}d^2x  g^{ab} \partial_a n^\mu \partial_b n^\mu, \nonumber 
\end{eqnarray} 
where $b[kT]$ is the bending rigidity \cite{HELFRICH-1973,POLYAKOV-NPB1986,KLEINERT-PLB1986}. The matrix $g_{ab}(a,b\!=\!1,2)$ in $S_1$ and $S_2$ is the metric on $M$, $g$ is the determinant of  $g_{ab}$, and $g^{ab}$ is its inverse. The symbol $n^\mu$ in $S_2$ is a unit normal vector of the surface. The continuous Hamiltonian is invariant under the conformal transformation such as $g_{ab}\to g_{ab}^\prime \!=\!fg_{ab}$ for any positive function $f$, and it is also invariant under the reparametrization such as $x\to x^\prime$, which changes both of the variables $X$ and $g$. These symmetries allows us to use a constant metric $g_{ab}=\delta_{ab}$, however, in the case of constant metric $g_{ab}=\delta_{ab}$ the in-plane deformation of $M$ is prohibited, and as a consequence the area of surface $X(M)$ remains constant as long as the surface tension coefficient $a$ in $S=aS_1\!+\!bS_2$ is fixed. No information of the in-plane transformation in membranes is obtained from such a model with $g_{ab}=\delta_{ab}$. 

The discrete Hamiltonians in Eqs. (\ref{Disc-Eneg})--(\ref{Disc-Eneg-S2}) are obtained from the continuous one in Eq. (\ref{cont_S}) as follows. On the triangulated surfaces shown in Figs. \ref{fig-1}(a)--\ref{fig-1}(c), the partial derivatives in $S_1$ in Eq. (\ref{cont_S}) can be replaced by $\partial_1 X^\mu \to X^\mu_2\!-\!X^\mu_1$, $\partial_2 X^\mu \to X^\mu_3\!-\!X^\mu_1$, 
where $X^\mu_i (\in {\bf R}^3)$ denotes the position of the vertex $i$ such that $\ell_1\!=\!|X^\mu_2\!-\!X^\mu_1|$, $\ell_2\!=\!|X^\mu_3\!-\!X^\mu_1|$. The derivatives in $S_2$ in Eq. (\ref{cont_S}) can also be replaced by $\partial_1 n^\mu \to {\bf n}_0\!-\!{\bf n}_2$, $\partial_2 n^\mu \to {\bf n}_0\!-\!{\bf n}_1$, where ${\bf n}_i(i\!=\!0,1,2,3)$ are shown in Fig. \ref{fig-1}(d).

We make $S_1$ and $S_2$ to be symmetric under the permutation of the indices of $\ell_i$ and ${\bf n}_i$ such that $1\to 2$, $2\to 3$, $3\to 1$. By including those terms which are cyclic under the permutation, and by multiplying a factor $1/3$, we have the discrete $S_1$ and $S_2$ in  Eqs. (\ref{Disc-Eneg})--(\ref{Disc-Eneg-S2}).

 This symmetrization is necessary, because if it were not for the symmetrization, the Hamiltonian becomes dependent on the choice of local coordinates on the triangles. Thus, the symmetrization $1\to 2$, $2\to 3$, $3\to 1$ of $S_1$ and $S_2$ is considered to be a lattice analogue of the reparametrization invariance of the continuous Hamiltonian in Eq. (\ref{cont_S}). We should note that the component $F(=\cos \Phi)$ of $g$ remains unchanged under the symmetrization. The reason of this is because the internal angle $\Phi$ is assumed to be independent of the vertices just as shown in Eq. (\ref{triangle-rel-2}). 

\section{Monte Carlo technique}
The Metropolis Monte Carlo simulation technique is employed to update the variables $X$, $g$ and $\rho$. The update of these variables is accepted with the probability ${\rm Min}[1,\exp(-\delta S)]$, where $\delta S\!=\!S({\rm new})\!-\!S({\rm old})$. 

The vertex position $X$ is updated such that $X\to X^\prime\!=\!X\!+\!\delta X$ with a random vector $\delta X$ in a small sphere. The function $F$ is updated such that $F^\prime\!=\!F+\delta F$, where $\delta F\left( \in [-0.5,0.5]\right)$ is a random number, under the constraint $|F^\prime|\!<\!1$. The density field $\rho$ is updated by  $\rho\to \rho^\prime\!=\!\rho+\delta \rho$, where $\delta \rho\left( \in [-0.5,0.5]\right)$. The constraint $\rho^\prime \!>\!0$ is imposed on this update. The acceptance rate for $X$ is approximately $60\%$, and those for $F$ and $\rho$ are $80\%$ and $90\%$ respectively.

One Monte Carlo sweep (MCS) consists of $N$ sequential updates of $X$, $N_T$ sequential updates of $F$, and $N_T$ sequential updates of $\rho$. The total number of MCS performed after sufficiently large number of thermalization MCS is $1.5\times 10^9 \sim 2\times 10^9$ at the transition region on the $N\!=\!8412$ and $N\!=\!12252$ surfaces, and relatively small number of MCS is performed at non-transition region and on smaller surfaces.  

\section{Simulation results}\label{results}
To see the dependence of physical quantities on the bending rigidity $b$, we fix the parameters $\lambda_F$, $\lambda_{\rho A}$ and $\lambda_\rho$ to the following values:
\begin{eqnarray} 
\label{constants}
 \lambda_F=0,\quad \lambda_{\rho A}=1,\quad \lambda_\rho=1 \qquad({\rm case\; 1}), \nonumber \\
 \lambda_F=2,\quad \lambda_{\rho A}=2,\quad \lambda_\rho=0 \qquad({\rm case\; 2}). 
\end{eqnarray} 
In case 1, the interaction term $S_F$ is neglected while the density term $S_{\rho A}$ and the interaction term  $S_\rho$ are included in the Hamiltonian. In case 2, both terms $S_F$ and $S_{\rho A}$ are included while $S_\rho$ is neglected.

\begin{figure}[hbt]
\centering
\includegraphics[width=12.5cm]{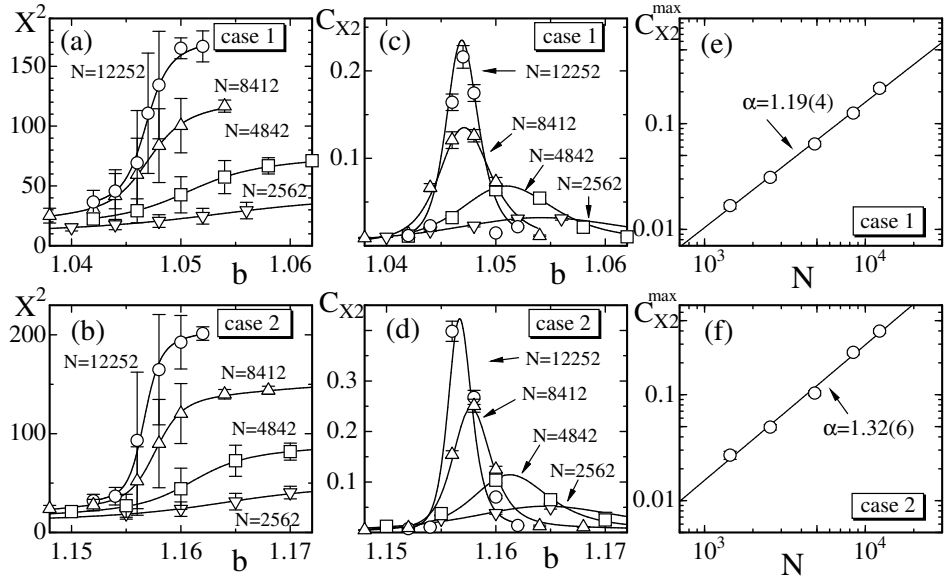}
\caption{The mean square size $X^2$ vs. $b$ in (a) case 1 and (b) case 2, the variance $C_{X^2}$ vs. $b$ in (c) case 1 and (d) case 2, and log-log plots of the peak $C_{X^2}^{\rm max}$ vs. $N$ in (e) case 1 and (f) case 2. The solid lines in (a), (b), (c) and (d) are drawn by the multi-histogram re-weighting technique. The error bars on the symbols denote the standard errors obtained by the binning analysis. }
\label{fig-2}
\end{figure}
We firstly show in Figs. \ref{fig-2}(a) and \ref{fig-2}(b) the mean square size 
\begin{equation}
\label{X2}
X^2={\frac 1 N} \sum_i \left(X_i-\bar X\right)^2, \quad \bar X={\frac 1 N} \sum_i X_i
\end{equation}
vs. $b$. The transition region $b$ in case 1 is slightly smaller than that in case 2, however, the behavior of $X^2$ against the variation $b$ in case 1 is almost identical to that in case 2. The error bars on $X^2$ is the standard errors obtained by the binning analysis, and the solid lines connecting the data symbols are drawn by the multi-histogram re-weighting technique \cite{Janke-histogram-2002}. The large errors shown at the transition region in Figs. \ref{fig-2}(a) and \ref{fig-2}(b) imply that $X^2$ discontinuously changes. 

Figures \ref{fig-2}(c) and \ref{fig-2}(d) show the variance $C_{X^2}$ of $X^2$ defined by
\begin{equation}
\label{CX2}
C_{X^2}={\frac 1 N} \langle \left(X^2-\langle X^2\rangle \right)^2 \rangle.
\end{equation}
We see that the peaks $C_{X^2}^{\rm max}$ increase with increasing $N$. To see the order of the transition, we show $C_{X^2}^{\rm max}$ vs. $N$ in Figs. \ref{fig-2}(e) and \ref{fig-2}(f) in log-log scales. The straight lines are drawn by a least squares fitting, and we have
\begin{eqnarray}
\label{CX2-scaling}
C_{X^2}^{\rm max}\sim N^{\alpha}, \quad
&&\alpha=1.19\pm 0.04\quad({\rm case\; 1}), \nonumber \\
&&\alpha=1.32\pm 0.06\quad({\rm case\; 2}). 
\end{eqnarray}
The results in Eq. (\ref{CX2-scaling}) indicate that $\alpha$ is slightly larger than $1$ in both cases. From the finite-size scaling theory  \cite{PRIVMAN-WS-1989,BINDER-RPP-1997,BNB-NPB-1993}, $\alpha\!=\! 1$ ($\alpha\!<\! 1$) implies that the transition is of first (second) order. Therefore, the results in Eq. (\ref{CX2-scaling}) imply that the model undergoes a discontinuous transition between the smooth spherical and collapsed phases just like the model without the metric degree of freedom \cite{Koibuchi-PRE2005}. 

\begin{figure}[hbt]
\centering
\includegraphics[width=12.5cm]{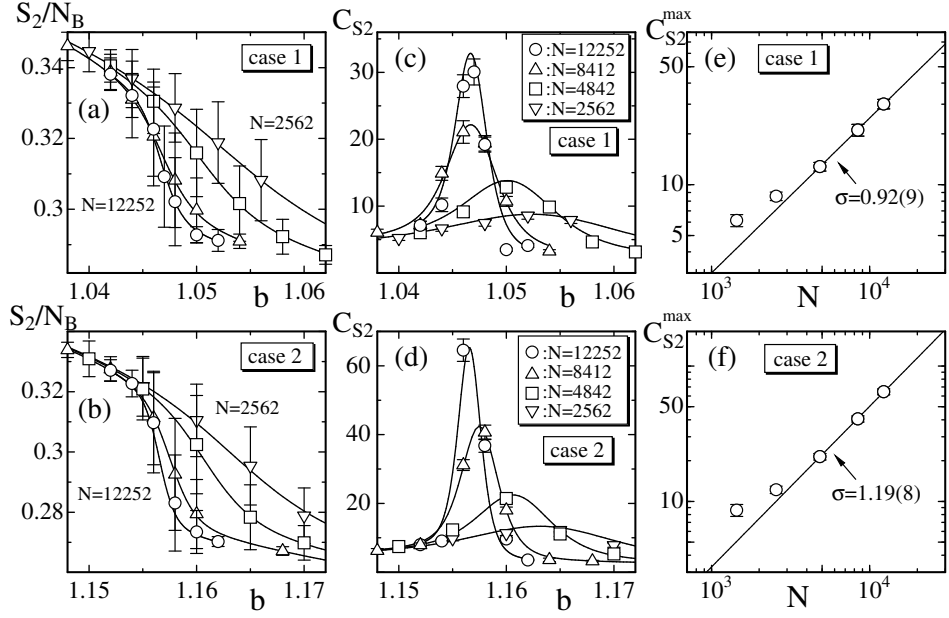}
\caption{The bending energy $S_2/N_B$ vs. $b$ in (a) case 1 and (b) case 2, the specific heat $C_{S_2}$ vs. $b$ in (c) case 1 and (d) case 2, and log-log plots of the peak $C_{S_2}^{\rm max}$ vs. $N$ in (e) case 1 and (f) case 2. The largest three data are used for the least squares fitting in both (e) and (f).}
\label{fig-3}
\end{figure}
 Large errors seen in $S_2/N_B$ also imply a discontinuous change in $S_2/N_B$  at the transition region (Figs. \ref{fig-3}(a), \ref{fig-3}(b)). The specific heat is defined by
\begin{equation}
\label{CS2}
C_{S_2}= {\frac {b^2}  N} \langle \left(S_2-\langle S_2 \rangle \right)^2 \rangle.
\end{equation}
 We see an expected peak $C_{S_2}^{\rm max}$ in each $C_{S_2}$ and find that $C_{S_2}^{\rm max}$ increases with increasing $N$ at the  transition point $b$ where $C_{X^2}$ has its peak. The $C_{S_2}^{\rm max}$ vs. $N$ are plotted in Figs. \ref{fig-3}(e) and \ref{fig-3}(f) in log-log scales. The straight lines  are drawn by  fitting the largest three data, and we have 
\begin{eqnarray}
\label{CS2-scaling}
C_{S_2}^{\rm max}\sim N^{\sigma}, \quad
&&\sigma=0.92\pm 0.09\quad({\rm case\; 1}), \nonumber \\
&&\sigma=1.19\pm 0.08\quad({\rm case\; 2}).
\end{eqnarray}
We find that both of the results in Eq. (\ref{CS2-scaling}) satisfy $\sigma\simeq 1$ and are consistent with the result of the conventional model in Ref. \refcite{Koibuchi-PRE2005}. This indicates that the model undergoes a first-order transition of surface fluctuations from the finite-size scaling theory \cite{PRIVMAN-WS-1989,BINDER-RPP-1997,BNB-NPB-1993}.   

\begin{figure}[htb]
\centering
\includegraphics[width=8.5cm]{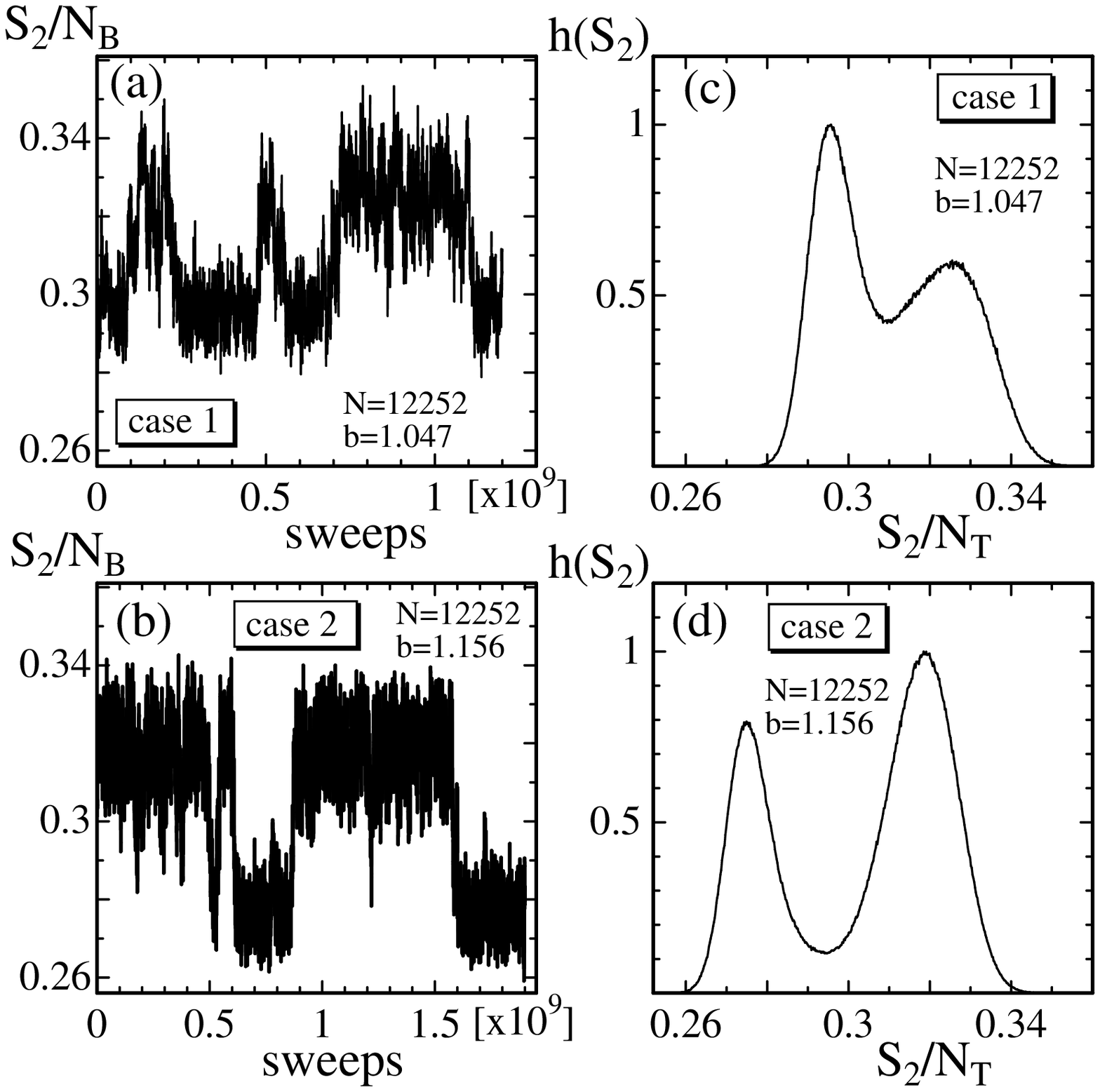}
\caption{The variation of $S_2/N_B$ against MCS in (a) case 1 and (b) case 2, and (c), (d) the normalized distribution (or histogram) $h(S_2)$ of $S_2/N_B$  corresponding to (a), (b), respectively.}
\label{fig-4}
\end{figure}
In order to see a discontinuity in $S_2/N_B$ more clearly, we show the variation of $S_2/N_B$ against MCS in Figs. \ref{fig-4}(a) and \ref{fig-4}(b) at the transition points. The corresponding distribution (or histogram) $h(S_2)$ of  $S_2/N_B$ are shown in Figs. \ref{fig-4}(a) and \ref{fig-4}(b). We find  a double peak structure in these $h(S_2)$. This structure depends on $N$ and becomes more apparent as $N$ increases, although the dependence of  $h(S_2)$ on $N$ is not shown in the figures. This implies that $S_2/N_B$ discontinuously changes at the transition point and indicates that the model undergoes a first-order transition in both cases. We should note that the transition occurs only four times during $1.9\times 10^9$ MCS at the transition point of the $N\!=\!12252$ surface (Fig. {\ref{fig-4}}(b)). This is the reason for the large errors in data $X^2$ and $S_2/N_B$ as mentioned above. Reliable estimate of critical exponents is very difficult at the first-order transition point.  
  
\begin{figure}[htb]
\centering
\includegraphics[width=8.5cm]{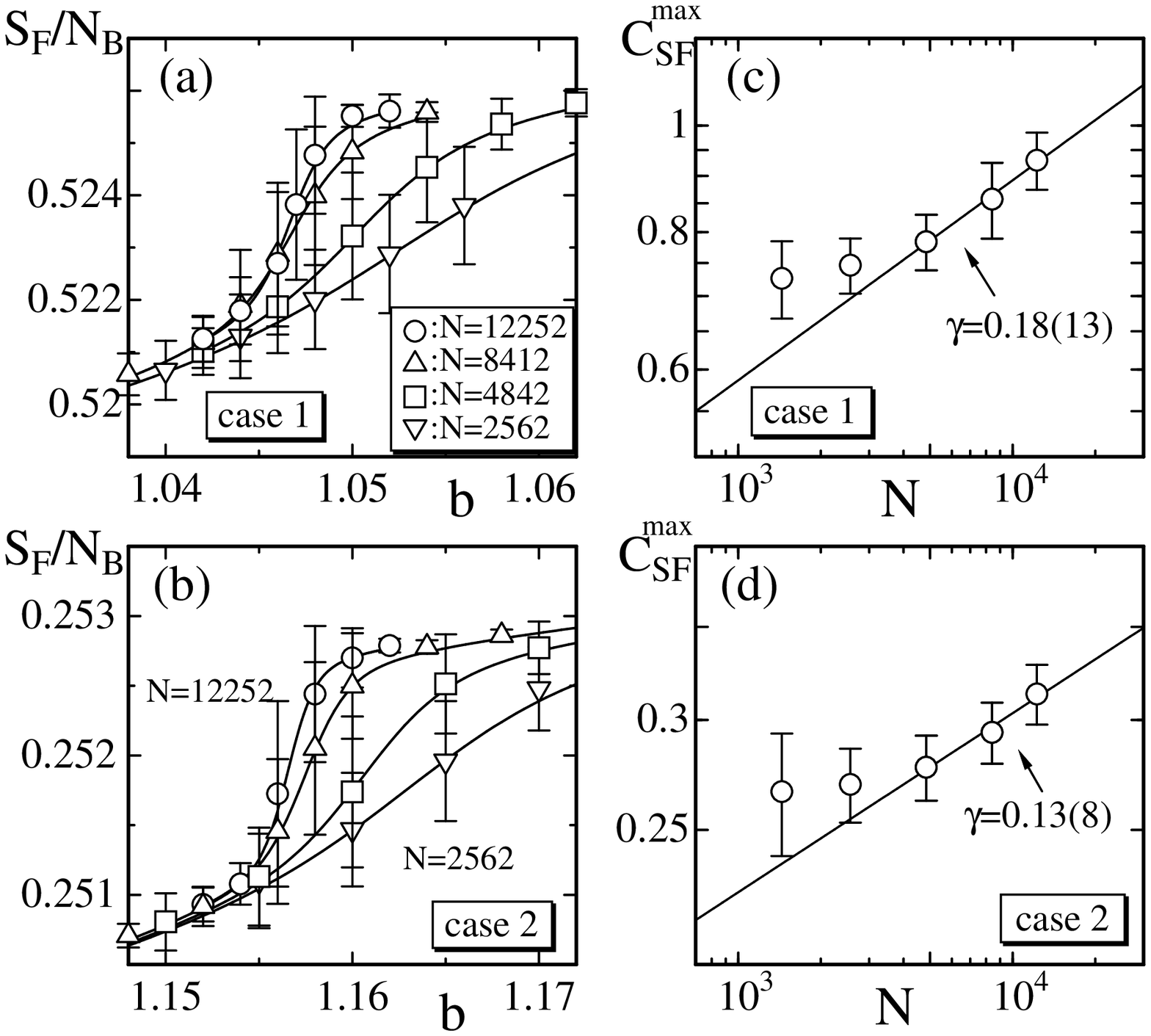}
\caption{The interaction energy $S_F/N_B$ vs. $b$ in (a) case 1 and (b) case 2, and log-log plots of $C_F^{\rm max}$ vs. $N$ in (c) case 1 and (d) case 2. The largest three data in (c) and (d) are fitted to Eq. (\ref{CFMAX}).}
\label{fig-5}
\end{figure}

The interaction energy $S_F/N_B$ vs. $b$ is plotted in Figs. \ref{fig-5}(a) and \ref{fig-5}(b). The peak values $C_F^{\rm max}$ of the variance $C_{S_F}\!=\!(1/N) \langle \left(S_F\!-\!\langle S_F\rangle \right)^2 \rangle$ are plotted against $N$ in Figs. \ref{fig-5}(c) and \ref{fig-5}(d) in a log-log scale. The largest three data are found to scale such that
\begin{eqnarray}
\label{CFMAX}
C_F^{\rm max}\sim N^\gamma, \quad &&\gamma=0.18\pm0.13\quad ({\rm case\; 1}), \nonumber \\
                                  &&\gamma=0.13\pm0.08\quad ({\rm case\; 2}).
\end{eqnarray}
These values satisfy $\gamma < 1$ and hence indicate that model undergoes a continuous transition of in-plane deformation \cite{PRIVMAN-WS-1989,BINDER-RPP-1997,BNB-NPB-1993}. The surface density $\rho$, which is not plotted in the figures, behaves just like $S_F/N_B$ in Figs. \ref{fig-5}(a) and \ref{fig-5}(b). However, we find no power low scaling behavior in $C_\rho^{\rm max}$ in contrast to $C_F^{\rm max}$ in Eq. (\ref{CFMAX}).

\begin{figure}[htb]
\centering
\includegraphics[width=12.5cm]{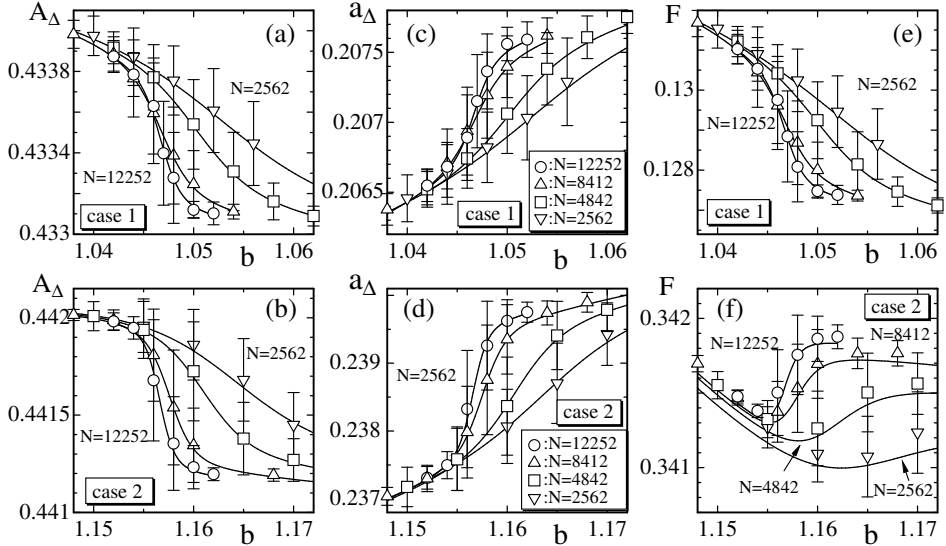}
\caption{The area $A_{\it \Delta}$ vs. $b$ in (a) case 1 and (b) case 2, the area $a_{\it \Delta}$ vs. $b$ in (c) case 1 and (d) case 2, and the variance the variable $F$ vs. $b$ in (e) case 1 and (f) case 2.}
\label{fig-6}
\end{figure}
The surface area defined by Eq. (\ref{area}) is an internal geometric variable, and hence it is interesting to see whether or not the phase transitions are reflected in  $A_{\it \Delta}$. Figures \ref{fig-6}(a) and \ref{fig-6}(b) show $A_{\it \Delta}(\!=\!\sum_{\it \Delta} A_{\it \Delta}/N_T)$ vs. $b$ in case 1 and case 2. We see that the variation of $A_{\it \Delta}$ against $b$ is quite analogous to that of $S_2/N_B$ shown in Fig. \ref{fig-3}, although we see that the change of $A_{\it \Delta}$ is very small compared to the value $A_{\it \Delta}$ itself. The variance  $C_{A_{\it \Delta}}$ defined by $C_{A_{\it \Delta}}\!=\!(1/N) \langle \left(\sum_{\it \Delta} A_{\it \Delta}\!-\!\langle \sum_{\it \Delta} A_{\it \Delta}\rangle \right)^2 \rangle$, which is not shown as a figure, is also analogous to $C_{X^2}$. However, the "internal" transition characterized by the fluctuations of $A_{\it \Delta}$ is not identified as a phase transition. Indeed, both of the data $C_{A_{\it \Delta}}^{\rm max}$ vs. $N$ do not satisfy the scaling relation $C_{A_{\it \Delta}}^{\rm max}\sim N^\mu$. Nevertheless,  the surface area $A_{\it \Delta}$ almost discontinuously changes against the variation of $b$ at the transition region (Figs. \ref{fig-6}(a),\ref{fig-6}(b)). 

We should note that the variation of the area $a_{\it \Delta}(\!=\!\sum_{\it \Delta} a_{\it \Delta}/N_T)$ of the surface $X(M)$ in Figs. \ref{fig-6}(c) and \ref{fig-6}(d) is also small compared to the value of $a_{\it \Delta}$ itself. However, $a_{\it \Delta}$ varies almost discontinuously just like $A_{\it \Delta}$. This implies that the transition of shape transformation is reflected in the in-plane surface deformation. 

The behavior of $F$ shown in Fig. \ref{fig-6}(e) is identical to that of $A_{\it \Delta}$ in Fig. \ref{fig-6}(a), while the behavior of $F$ vs. $b$ in  Fig. \ref{fig-6}(f) is different from  that of $A_{\it \Delta}$ in Fig. \ref{fig-6}(b). We consider that the interaction described by $S_F$ in Eq. (\ref{Disc-Eneg-Sg}) is an origin of this difference. Thus, we find that the in-plane phase structure is sensitive to the measure factor in Eq. (\ref{measure}) such as the factor of $S_F$. 

\begin{figure}[htb]
\centering
\includegraphics[width=9.5cm]{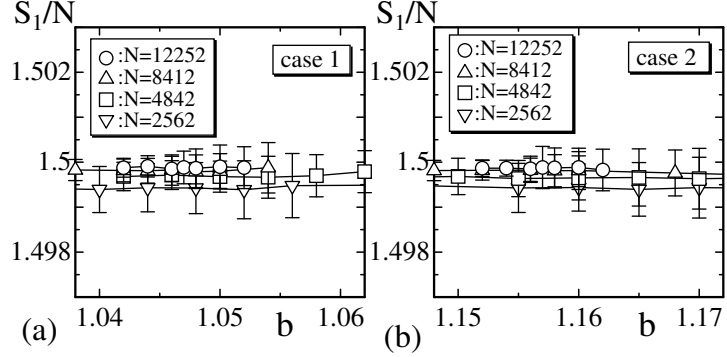}
\caption{The Gaussian bond potential $S_1/N$ vs. $b$ in (a) case 1 and (b) case 2. The solid lines are drawn as a guide to the eyes.}
\label{fig-7}
\end{figure}
Finally in this section, we show the Gaussian bond potential $S_1/N$ vs. $b$ in Figs. \ref{fig-7}(a) and \ref{fig-7}(b). We see that the expected relation $S_1/N\!\to\! 3/2$ $(N\!\to\! \infty)$ is satisfied in both cases. This implies that the simulations are successfully performed. We should emphasize that in both cases $S_1$ is different from either $A_{\it \Delta}$ or $a_{\it \Delta}$, and hence $A_{\it \Delta}$ and $a_{\it \Delta}$ are not always constrained to be a constant. This is in sharp contrast to the case of the conventional model,  where the Gaussian bond potential $\sum_{ij}(X_i\!-\!X_j)^2$  remains constant and corresponds to the surface area. 

\section{Summary and Conclusion}
In this paper, using the canonical Monte Carlo simulation technique, we have studied a triangulated surface model, in which the metric is assumed as a dynamical variable. A deficit angle for triangles of triangulated surface is assumed as the metric degrees of freedom.  We focus on whether or not the collapsing transition accompanies an in-plane surface deformation. It is also interesting to see whether or not the phase structure of the model is identical to the one of the model with a constant metric such as the Euclidean metric {$g_{ab}\!=\!\delta_{ab}$} or the induced metric {$g_{ab}\!=\!\partial_a X^\mu \partial_b X^\mu$} of the mapping {$X:M\to {\bf R}^3$}.

We find that the model undergoes a first-order transition between the smooth spherical phase at $b\!\to\! \infty$ and the collapsed phase at $b\!\to\! 0$. The transition is almost identical to the one observed in the model with constant metric in Refs. \refcite{Koibuchi-NPB2006,Koibuchi-PRE2005}. This  indicates that the transition of shape transformations is not influenced by the metric variable in Eq. (\ref{induced_metric}). We also find that the phase transition accompanies a continuous in-plane transition, which is confirmed by the finite-size scaling analyses for the peak values of the variance $C_{S_F}$. This in-plane transition is reflected in an almost discontinuous change in both of the internal and external quantities such as the surface density and the external surface area, although the discontinuous changes are very small compared to the values themselves.  

\vspace{0.5cm}
\noindent
{\bf Acknowledgment}
This work is supported in part by a Promotion of Joint Research of Toyohashi University of Technology.





\end{document}